\documentstyle[12pt,epsfig,cite]{article}
\newcommand{\be}{\begin{equation}}
\newcommand{\ee}{\end{equation}}
\newcommand{\bea}{\begin{eqnarray}}
\newcommand{\eea}{\end{eqnarray}}

\newcommand{\la}{\langle}
\newcommand{\ra}{\rangle}

\newcommand{\p}{\partial}

\newcommand{\SL}[0]{SL(2,R)}
\newcommand{\lr}{\leftrightarrow}

\def\CP{{\mathcal{P}}}
\def\CQ{{\mathcal{Q}}}
\def\IB{\relax\hbox{$\inbar\kern-.3em{\rm B}$}}
\def\IC{\relax\hbox{$\inbar\kern-.3em{\rm C}$}}
\def\ID{\relax\hbox{$\inbar\kern-.3em{\rm D}$}}
\def\IE{\relax\hbox{$\inbar\kern-.3em{\rm E}$}}
\def\IF{\relax\hbox{$\inbar\kern-.3em{\rm F}$}}
\def\IG{\relax\hbox{$\inbar\kern-.3em{\rm G}$}}
\def\IGa{\relax\hbox{${\rm I}\kern-.18em\Gamma$}}
\def\IH{\relax{\rm I\kern-.18em H}}
\def\IK{\relax{\rm I\kern-.18em K}}
\def\IL{\relax{\rm I\kern-.18em L}}
\def\IP{\relax{\rm I\kern-.18em P}}
\def\IR{\relax{\rm I\kern-.18em R}}
\def\IZ{\relax{\rm Z\kern-.5em Z}}

\def\half{\frac{1}{2}}
\def\p{\partial}

\begin{document}

\begin{titlepage}

\rightline{OUTP-00-30-P}
\rightline{July 2000}

\vskip 2 cm

\begin{center}
{\LARGE Logarithmic operators in the $\SL$ WZNW model}
\vskip 1 cm

{\large A. Nichols\footnote{a.nichols1@physics.ox.ac.uk} and Sanjay\footnote{s.sanjay1@physics.ox.ac.uk}}

\vskip 1.5 cm

{\it Theoretical Physics, Department of Physics, University of Oxford}\\
{\it 1 Keble Road, Oxford, OX1 3NP,  UK}

\end{center}

\begin{abstract}
We find some exact solutions of the Knizhnik-Zamolodchikov equation for the four point correlation functions that occur in the $\SL$ WZNW model. They exhibit logarithmic behaviour in both the Kac-Moody and Virasoro parts. We discuss their implication for the operator product expansion. We also observe the appearance of several symmetries of the correlation functions. 

\end{abstract}

\end{titlepage}

\newpage

\section{Introduction}

The $\SL$ WZNW model is an old and interesting problem. The case of WZNW models based on a compact Lie group such as $SU(2)$ was completely analysed in \cite{Knizhnik:1984nr,Gepner:1986wi}. In this case the unitary representations are finite dimensional and correlation functions can be written as a finite sum of conformal blocks by the bootstrap approach. The non-compact groups gives a much more complicated situation and the general solution is not known. There has been much work done on analysing this \cite{Zamolodchikov:1986bd,PGB}.

The group $\SL$, which is one of the simplest non-compact groups, is important in many key areas. 

The first is in two dimensional gravity \cite{Polyakov:1987zb,Knizhnik:1988ak,Dorn:1994xn, Zamolodchikov:1996aa}. The gravitational Ward identities for correlation functions of the metric are precisely the same as the $\SL$ KZ equations \cite{Polyakov:1987zb,Knizhnik:1988ak}. If matter is coupled to gravity then the gravitationally dressed matter correlation functions also obey identities of the same form as $\SL$ KZ equations \cite{Klebanov:1993ms,Bilal:1994nx}. The case of string theory on $AdS_3$ is described by an exact conformal field theory - the $\SL$ WZNW theory. There has recently been much renewed interest in this due to the AdS/CFT correspondence \cite{Maldacena:1998re,Gubser:1998bc,Witten:1998qj} although so far correlation functions have only been computed in the leading supergravity approximation. The final application is with respect to the Quantum Hall plateau transition where the $\SL$ WZNW model has recently been proposed as a low energy effective  field theory \cite{Bhaseen:1999nm,Kogan:1999hz} (See also \cite{Zirnbauer:1999ua} where a more general $PSL(2|2)$ group was discussed, and earlier references \cite{KCC}.)

Correlation functions in WZNW model satisfy Knizhnik-Zamolodchikov (KZ) equation \cite{Knizhnik:1984nr} which is a set of coupled partial differential equations coming from constraints due to  null states. For compact groups like $SU(N)$ it can be solved exactly for four point functions by the bootstrap approach which reduces it to a set of ordinary differential equations. This is due to the fact that only a finite number of conformal blocks occur. In the $\SL$ case one particular four point function was found \cite{Kogan:1999hz} that had a logarithmic behaviour. In this paper we find some more exact solutions of KZ equation which are logarithmic in nature and show that this cannot be understood if the only fields present are primary operators.

These logarithmic singularities in four point functions have a natural interpretation in terms of logarithmic conformal field theory (LCFT) \cite{Gurarie:1993xq}. Correlation functions in LCFT have a special form. In particular operator product expansion (OPE) of two operators can receive contribution from logarithmic operators  whenever the dimensions of the two or more operators become degenerate. This gives logarithms of cross ratios in four point functions. LCFTs are also discussed earlier in connection with the WZNW model \cite{Caux:1997kq,Kogan:1997fd}. The presence of logarithmic singularities in the four point functions in $\SL$ was discussed in \cite{Bilal:1994nx,Kogan:1997vx}  but only in some asymptotic limits.

The plan of the paper is as follows. In next section we discuss correlation functions in the $\SL$ WZNW model. In the following section we present some exact solutions of KZ equation and also some asymptotic solutions which are logarithmic in nature. Finally we make some comments about the OPE structure in $\SL$ WZNW theory and the appearance of several symmetries.

\section{Correlation functions}

The WZNW theory has affine $\SL$ left and right moving symmetries whose modes generate the Kac-Moody algebra:
\bea
~[ J^3_n,J^3_m ] & = & -\frac{k}{2}n \delta_{n+m,0} \nonumber \\
~[ J^3_n,J^{\pm}_m ] & = & {\pm} J^{\pm}_{n+m} \\
~[ J^+_n,J^-_m ] & = & -2J^3_{n+m}+kn \delta_{n+m,0} \nonumber
\eea
We have similar expressions for the $\bar{J}^a_n$. The zero mode algebra is just the group $\SL$. We introduce the following representation for the  $\SL$ generators \cite{Zamolodchikov:1986bd}:
\bea \label{eq:repn}
J^+=-x^2\frac{\p}{\p x}-2jx, ~~~ 
J^-=-\frac{\p}{\p x}, ~~~
J^3=x\frac{\p}{\p x}+j \eea
There is also a similar algebra in terms of $\bar{x}$ for the antiholomorphic part. From now on we drop the antiholomorphic dependence in our notation although it can easily be restored. They obey the standard relations:
\bea \label{eq:SLalgeb} [J^+,J^-]=-2J^3 ~~~~
[J^3,J^{\pm}]=\pm J^{\pm} 
\eea

The quadratic Casimir is:
\bea C_2=\eta_{ab} J^a J^b=\half J^+J^- + \half J^-J^+ -J^3J^3=-j(j-1) \eea

The stress-energy tensor in the case of the ungauged WZNW model is given in terms of the normal Sugawara construction:

\be
T(z)=\frac{1}{k-2}\eta_{ab} :J^a(z) J^b(z):
\ee

where : : denotes normal ordering. In terms of a mode expansion:

\be
\label{eq:Sugawara}
T(z)=\sum_n{\frac{L_n}{z^{n+2}}} \quad L_n=\frac{1}{k-2}\sum_m{\eta_{ab}:J^a_m J^b_{n-m}:}
\ee

The modes $L_n$ obey the standard Virasoro algebra with central charge $c=\frac{3k}{k-2}$.

We introduce primary fields, $\phi_j(x,z)$  of the KM algebra:

\be
J^a(z)\phi_j(x,w) = \frac{1}{z-w} J^a \phi_j(x,w) 
\ee

where $J^a$ is given by (\ref{eq:repn}). The fields $\phi_j(x,z)$ are also primary with respect to the Virasoro algebra with $L_0$ eigenvalue:

\be
h=-\frac{j(j-1)}{k-2}
\ee

The two point functions are fully determined using global $\SL$ and conformal transformations and can be normalised in the standard way:

\be
\la \phi_{j_1}(x_1,z_1) \phi_{j_2}(x_2,z_2) \ra = \delta^{j_1}_{j_2}x_{12}^{-2j_1}z_{12}^{-2h}
\ee

The general form of the three point function is:

\bea
\la \phi_{j_1}(x_1,z_1) \phi_{j_2}(x_2,z_2) \phi_{j_3}(x_3,z_3) \ra = C(j_1,j_2,j_3)~~ x_{12}^{-j_1-j_2+j_3} x_{13}^{-j_1-j_3+j_2} x_{23}^{-j_2-j_3+j_1} \\
z_{12}^{-h_1-h_2+h_3} z_{13}^{-h_1-h_3+h_2} z_{23}^{-h_2-h_3+h_1} \nonumber
\eea

The $C(j_1,j_2,j_3)$ are the structure constants which in principle completely determine the entire theory. For the case of $\SL$ these were found in \cite{otherapp}. In the case of the closely related Liouville theory these were found in \cite{Dorn:1994xn,Zamolodchikov:1996aa}.

For the case of the four point correlation functions of $\SL$ primaries the form is determined by global conformal and $\SL$ transformations up to a function of the cross ratios.

\bea \label{eq:correl} 
\langle \phi_{j_1}(x_1,z_1) \phi_{j_2}(x_2,z_2) \phi_{j_3}(x_3,z_3) \phi_{j_4}(x_4,z_4) \rangle
&=&z_{43}^{h_2+h_1-h_4-h_3}z_{42}^{-2h_2}z_{41}^{h_3+h_2-h_4-h_1} \nonumber \\
& & z_{31}^{h_4-h_1-h_2-h_3}x_{43}^{j_2+j_1-j_4-j_3}x_{42}^{-2j_2} \nonumber \\
& & x_{41}^{j_3+j_2-j_4-j_1}x_{31}^{j_4-j_1-j_2-j_3}~F(x,z)
\eea 

Here the invariant cross ratios are:
\be 
x=\frac{x_{21}x_{43}}{x_{31}x_{42}} ~~~ z=\frac{z_{21}z_{43}}{z_{31}z_{42}} 
\ee

In a normal CFT we expect the OPE of two primary fields to take the general form:

\bea
\label{eq:OPE}
\phi_{j_1}(x_1,z_1) \phi_{j_2}(x_2,z_2) = \sum_J{C(j_1,j_2,J) z_{12}^{-h_1-h_2+h_J} x_{12}^{-j_1-j_2+J} [ \phi_J(x_2,z_2) ] } 
\eea

where we have by denoted $[ \phi_J ]$ all fields that can be produced from the given primary field $ \phi_j $. In principle given $C(j_1,j_2,j_3)$, we know the entire operator content of the theory and should be able to determine all higher point correlation functions using the OPE (\ref{eq:OPE}) and the crossing symmetries. This is called conformal bootstrap and has only so far been solved for the minimal models. We will see however that our solutions require more operators to be included in the OPE.

\section{The Knizhnik-Zamolodchikov Equation}
Correlation functions of the WZNW model satisfy a set of partial differential equations known as  Knizhnik-Zamolodchikov (KZ) equation due to constraints from the null states following from (\ref{eq:Sugawara}). These are:

\be
|\chi \ra = ( L_{-1} - \frac{1}{k-2} \eta_{ab}J^a_{-1}J^b_0 ) |\phi \ra
\ee

For two and three point functions this gives no new information. However for the four point function (\ref{eq:correl}) it becomes a partial differential equation for $F(x,z)$. For  a compact Lie group this equation can be solved \cite {Knizhnik:1984nr} as it reduces to a set of ordinary differential equations. For a non-compact group the situation is much more complicated. Below we specialise to the case of the group $\SL$ in which case the KZ equation is similar to that for $SU(2)$ which was considered in \cite{Zamolodchikov:1986bd}:

\be
\left[(k-2) \frac{\p}{\p z_i}+\sum_{j\neq i}\frac{\eta_{ab} J^a_i \otimes J^b_j}{z_i-z_j} \right] \left<\phi_{j_1}(z_1) \cdots \phi_{j_n}(z_n) \right> =0 
\ee

where $k$ is the level of $\SL$ WZNW model.

If we now use our representation (\ref{eq:repn}) we find the KZ equation for four point functions is:

\be \label{eq:KZ}
(k-2) \frac{\p}{\p z} F(x,z)=\left[ \frac{\CP}{z}+\frac{\CQ}{z-1} \right] F(x,z)
\ee

Explicitly these are:

\bea
\CP \!\!\!\!&=&\!\!-x^2(1-x)\frac{\p^2}{\p x^2}+((j_1+j_2+j_3-j_4+1)x^2-2j_1x-2j_2x(1-x))\frac{\p}{\p x} \nonumber \\
& & +2j_2(j_1+j_2+j_3-j_4)x-2j_1j_2 \\
\CQ \!\!\!\!&=&\!\!-(1-x)^2x\frac{\p^2}{\p x^2}-((j_1+j_2+j_3-j_4+1)(1-x)^2-2j_3(1-x)-2j_2x(1-x))\frac{\p}{\p x} \nonumber \\
& & +2j_2(j_1+j_2+j_3-j_4)(1-x)-2j_2j_3 
\eea

These are the same expressions obtained in \cite{Teschner:1999ug} except we differ in notation by $j \rightarrow -j$.

\section{Solutions of the KZ equation}

\subsection{Exact solutions}

We seek solutions of the KZ equation which are of the following form:
\be 
F(x,z)=z^{\frac{\alpha}{k-2}}(1-z)^{\frac{\beta}{k-2}}x^p(1-x)^q[X(x)+Z(z)]
\ee
This ansatz allows us to see the various crossing symmetries in the solution nicely. Here we summarize the solutions leaving the details to the Appendix. Also we restrict ourselves to real values of $j$ (Discrete series; see Appendix for details about $\SL$ representation theory) in this section and in the rest of the paper. 

\begin{itemize}

\item $j_1=j_2=j_3=j_4=0$ This is the solution found previously in \cite{Kogan:1999hz}.
\be
F(x,z)=A\left[\frac{\ln z }{k-2}-\ln x\right] + B\left[\frac{\ln (1-z)}{k-2}-\ln (1-x) \right]+C
\ee

\item Logarithms in $x$ and $z$.
\begin{enumerate}

\item $j_1=j_2=0 ~~ j \equiv j_3=j_4 \neq \half$.
\be
F(x,z)=A \left[ \frac{\ln(z-1)}{k-2}+\frac{\ln(x-1)}{2j-1} \right]+B
\ee
This obeys the crossing symmetry $1 \lr 2 \Rightarrow x \lr \frac{x}{x-1} , z \lr \frac{z}{z-1}$. 

\item $j_1=j_3=0 ~~ j \equiv j_2=j_4 \neq \half$.
\be
F(x,z)=A \left[\frac{\ln z-\ln(z-1)}{k-2}+\frac{\ln x-\ln(x-1)}{2j-1}\right]+B
\ee
There is now no reason to expect symmetry under $1 \lr 2$ but instead we have $1 \lr 3 \Rightarrow x \lr 1-x,z \lr 1-z$. This solution is also well behaved as $x,z\rightarrow \infty$

\item $j_1=j_4=0 ~~ j \equiv j_2=j_3 \neq \half$.
\be
F(x,z)=A \left[\frac{\ln z}{k-2}+\frac{\ln x}{2j-1} \right]+B
\ee
Here we see the symmetry $1 \lr 4 \Rightarrow x \lr \frac{1}{x},z \lr \frac{1}{z}$.

\end{enumerate}

\item $j_2=0 ~~ j_1+j_3+j_4-1=0$ ~~~ Logarithms in $z$ disappear.

\bea
\langle \phi_{1-j_3-j_4}(x_1,z_1) \phi_{0}(x_2,z_2) \phi_{j_3}(x_3,z_3) \phi_{j_4}(x_4,z_4) \rangle=
z_{43}^{h_1-h_4-h_3}z_{41}^{h_3-h_4-h_1} \nonumber \\
z_{31}^{h_4-h_1-h_3}x_{43}^{1-2j_3-2j_4} \nonumber \\
z_{41}^{2j_3-1}x_{31}^{2j_4-1}~\int (1-x)^{-2j_3}x^{-2j_1} dx
\eea
This has logarithms if $j_1$ or $j_3 \in Z$.

\end{itemize}

There are also solutions related to the above by various symmetries which we mention later.

\subsection{Asymptotic solutions}

In \cite{Bilal:1994nx,Kogan:1997vx} some asymptotic solutions were found by considering the $x\rightarrow1$ asymptotics. However it was not clear, without having some exact solutions, that these persisted to all orders and that the logarithmic behaviour was genuine.
Here we look at the $z \rightarrow 0$ leading asymptotics. If we assume a power series in $z$ then the leading form of $F(x,z)$ obeys:

\be (k-2)\frac{\p F(x,z)}{\p z}=\frac{\CP F(x,z)}{z} \ee

We can easily solve this by separation of variables to get solutions:

\be
F_{\alpha}(x,z)=z^{\frac{\alpha}{k-2}}G_{\alpha}(x)
\ee
with:
\bea
G_{\alpha}(x)&=&A _2F_1(j_2-j_1+\half+\frac{\eta}{2},j_3-j_4+\half+\frac{\eta}{2};1+\eta;x)x^{-j_1-j_2+\half+\frac{\eta}{2}} \nonumber \\
&   &+B _2F_1(j_2-j_1+\half-\frac{\eta}{2},j_3-j_4+\half-\frac{\eta}{2};1-\eta;x)x^{-j_1-j_2+\half-\frac{\eta}{2}} \\
\eta&=&\sqrt{4{j_1}^2-4j_1+4{j_2}^2-4j_2+1-4\alpha} \nonumber
\eea

The value of $\alpha$ is arbitrary so far but if we wish to have agreement between the OPE and the tensor product of two representations then we should have $\alpha=0$ as this leads to:

\be
\phi_{j_1}(x,z)\phi_{j_2}(0,0)= \sum_{J=j_1+j_2} |z|^{-h_1-h_2+h_J}|x|^{-j_1-j_2+J}\left[ O_J(0,0) \right]
\ee
where $[O_J]$ denotes the conformal family of operators, which are not primaries in general, that have conformal dimension $J$.

We thus generically have logarithms in the Kac-Moody states if $\eta \in Z$.

\section{Structure of the OPE}
\subsection{Logarithmic Conformal Field Theory}

The occurrence of logarithmic singularities in conformal blocks signals the appearance of logarithmic operators. In a conformally invariant theory the OPE of primary operators is of the form 

\be
O_{i}(z_{1})O_{j}(z_{2})=\sum_{k}\frac{f_{ij}^{k}}{|z_{12}|^{h_{i}+h_{j}-h_{k}}}O_{k}(z_{1}) + ...
\ee

Two and three point functions are fixed by global conformal invariance up to a numerical constant. Four point functions are also fixed up to an arbitrary function of conformally invariant cross-ratios.

\be
<O_{i}(z_{1})O_{j}(z_{2})O_{k}(z_{3})O_{l}(z_{4})>
=\sum_{m,n}\frac{f_{ij}^{m}f_{kl}^{n}}{|z_{12}|^{h_{i}+h_{j}}|z_{34}|^{h_{k}+h_{l}}}F(z)
\ee

where $z$ is the conformally invariant cross-ratio defined as above. If the OPE is of the above form then $F(z)$ is a power series in $z$ and there can be no logarithmic singularities in four point functions. However we have explicit solutions which are logarithmic in nature. In that case it becomes necessary to modify the OPE and include the logarithmic operators:

\be
O_{i}(z_{1})O_{j}(z_{2})=\frac{1}{|z_{12}|^{h_{i}+h_{j}-h}}(C \ln|z_{12}|^{2} + D ) + ...
\ee

where operators $C$ and $D$ form a logarithmic pair with the degenerate dimension $h$. 
We have the following form of two point functions for $C$ and $D$ \cite{Gurarie:1993xq,Caux:1996nm}

\bea
\label{eq:logcorrel}
<C(z_{1})C(z_{2})>=0 \nonumber \\
<C(z_{1})D(z_{2})>=\frac{1}{|z_{12}|^{2h}} \\
<D(z_{1})D(z_{2})>=\frac{-2\ln|z_{12}|^2+ \delta}{|z_{12}|^{2h}} \nonumber
\eea

They have the following form of OPE with the stress-tensor $T(z)$:

\be
T(w)C(z)=\frac{h}{(w-z)^{2}}C(z)  +  \frac{1}{(w-z)}\frac{\p C}{\p z} + ...
\ee

\be
T(w)D(z)=\frac{h}{(w-z)^{2}}D(z)  + \frac{1}{(w-z)^{2}}C(z)+ \frac{1}{(w-z)}\frac{\p D}{\p z} + ...
\ee

Or equivalently:

\be
L_{0}|C>=h |C>  \quad   L_{0}|D>=h |D> + |C>
\ee

In this case $L_0$ cannot be diagonalised and $C,D$ are degenerate. It is immediately apparent, from the form of the two point functions, that such theories cannot be unitary as they possess extra zero norm states which do not decouple from the physical spectrum. The hamiltonian $L_{0}$ forms an indecomposable Virasoro module. These extra null states can reproduce the structure of a full gauge theory despite the fact we only have scalars involved \cite{Fronsdal:1989pp,Caux:1996nm} due to the symmetry of the OPE under $D \rightarrow D + \lambda C $. In \cite{Fronsdal:1989pp} this was discussed from the point of view of singletons. The relation between singletons and LCFT will be discussed later.

Correlation functions with logarithmic behaviour were found in the $GL(1,1)$ WZNW model in \cite{Rozansky:1993td}. There has been a lot of work on LCFTs \cite{LCFTrefs}, in particular the role of extended algebras in the general $c_{p,q}$ models and for the special $c=-2$ case. The indecomposable representations of the Virasoro algebra have also been investigated \cite{Rohsiepe:1996qj}.
As with a general CFT these are only well defined if correlators obey crossing symmetry and single-valuedness as well as modular invariance on the torus. This has been shown to hold for some classes of LCFTs so far.

We can also get a similar structure arising if we have logarithms in $x$. In this case it is $J_0^3$ that is of Jordan normal form \cite{Maassarani:1996jn,Kogan:1997vx,Kogan:1997fd}:

\be
J_0^3|C>=m |C>  \quad   J_0^3|D>=m |D> + |C>
\ee

In a more general case if we have degeneracy with the descendants of another primary then we will also have logarithms in the OPE.

We now wish to show how this general structure arises from our solutions.

\subsection{Exact Logarithmic Solutions and OPE}

\begin{itemize}
\item{$j=j_1=j_3 ~~~ j_2=j_4=0$}
\bea
\la \phi_j(x_1,z_1)\phi_0(x_2,z_2)\phi_{j}(x_3,z_3)\phi_{0}(x_4,z_4) \ra =|z_{31}|^{-2h} 
|x_{31}|^{-2j}~\biggl[ A\Bigl( \ln \left| \frac{1-x}{x} \right| + \\
\frac{2j-1}{k-2}\ln \left| \frac{1-z}{z} \right| \Bigr) + B  \biggr] \nonumber
\eea

If we take the limit $1 \rightarrow 3$ and $2 \rightarrow 4$ which corresponds to $x,z \rightarrow \infty$ we can expand our solution:

\be
\la \phi_j(x_1,z_1)\phi_0(x_2,z_2)\phi_{j}(x_3,z_3)\phi_{0}(x_4,z_4) \ra =|z_{31}|^{-2h}
|x_{31}|^{-2j}~\left[ 1 + ... \right]
\ee

where ... stands for subleading terms that vanish in the above limit. Then we find that:
\bea
\phi_j (x_1, z_1) \phi_j(x_3,z_3) &=&|z_{31}|^{-2h}|x_{31}|^{-2j}~ \left [1 + ... \right] \\
\phi_0 (x_2, z_2) \phi_0(x_4,z_4) &=& 1 + ...  
\eea

We also find the correct two point functions if we take the correlation function of the above. Note we can always rescale operators so that the leading contribution has coefficient unity. The proper coefficients can only be obtained by consistency with higher order terms in the expansions. We do not attempt this here. 

Now taking the  $1 \rightarrow 2$ and $3 \rightarrow 4$ limit which corresponds to $x,z \rightarrow 0$ we can again expand our solution:

\bea
\la \phi_j(x_1,z_1)\phi_0(x_2,z_2)\phi_{j}(x_3,z_3)\phi_{0}(x_4,z_4) \ra =|z_{42}|^{-2h}
|x_{42}|^{-2j}~\Bigl[ \ln|x_{21}| + \ln|x_{43}|   \nonumber \\
-2\ln|x_{42}| + \frac{2j-1}{k-2} \left(\ln|z_{21}| +\ln|z_{43}| - 2\ln|z_{42}| \right) ... \Bigr]
\eea

where we use $\ln x=\ln x_{21}+\ln x_{43} - \ln x_{31} -\ln x_{42}=\ln x_{21} + \ln x_{43} - 2 \ln x_{42} + ...$ and similarly for $z$. Again the subleading terms are ignored.
We now find that:

\bea
\phi_j (x_1, z_1) \phi_0(x_2,z_2) \!\!\! &=& \!\!E_j^x(x_2,z_2)\ln|x_{12}|+E_j^z(x_2,z_2)\ln|z_{12}| +F_j(x_2,z_2) + ...  \\
\phi_j (x_3, z_3) \phi_0(x_4,z_4) \!\!\! &=& \!\!E_j^x(x_4,z_4)\ln|x_{34}|+E_j^z(x_4,z_4)\ln|z_{34}| +F_j(x_4,z_4) + ... \nonumber
\eea

Using our exact solution leads to the following 2-pt functions when $j\ne 0$:
\bea
\la E_j^a E_j^b \ra = 0, \quad 
\la  E_j^z(x,z)F_j(0,0)\ra = \frac{2j-1}{(k-2)}|z|^{-2h}|x|^{-2j}  \\ \nonumber
\la E_j^x(x,z)F_j(0,0)\ra = |z|^{-2h}|x|^{-2j} \\
\la  F_j(x,z)F_j(0,0)\ra = -2|z|^{-2h}|x|^{-2j} \left(\frac{2j-1}{k-2}\ln|z|+\ln|x| \right) \nonumber
\eea

The fields $E_j^z,F_j$ form a Virasoro logarithmic pair \cite{Gurarie:1993xq} which mix under the conformal transformations $z \rightarrow \lambda z$:
\be
F_j \rightarrow F_j - \ln \lambda E_j^z, \quad
E_j^z \rightarrow E_j^z
\ee

$E_j^x,F_j$ form a Kac-Moody logarithmic pair \cite{Kogan:1997vx,Kogan:1997fd} which mix under the $\SL$ transformations $x \rightarrow \epsilon x$:
\be
F_j \rightarrow F_j-\ln \epsilon E_j^x, \quad
E_j^x \rightarrow E_j^x
\ee

If $j=\half$ it is consistent at this level to set $E_j^z=0$.

\item{$j_2=\half ~~~ j_1+j_3+j_4=\half$}

We find
\bea
\label{eq:nojzero}
F(x,z)=|x|^{-2j_1}|1-x|^{-2j_3}|z|^{-j_1}|1-z|^{-2j_3} \left| \int{x^{2j_1-1}(x-1)^{2j_3-1} dx} \right|
\eea
The correlation function is then obtained using (\ref{eq:correl}).

\underline{$j_1 \ne 0$}
\bea
\phi_{j_1}(x,z)\phi_{\half}(0,0)&=&|z|^{-h_1-h_2+h_J}|x|^{-j_1-j_2+J}O_J(0,0)+... \\
\phi_{j_3}(x,z)\phi_{j_4}(0,0)&=&|z|^{-h_3-h_4+h_J}|x|^{-j_3-j_4+J}O^{'}_J(0,0)+... \nonumber
\eea

where $J=j_1+\half$. Then:
\be
\la O_J(x,z)O^{'}_J(0,0) \ra = |z|^{-2h_J}|x|^{-2J}
\ee

\underline{$j_1 = 0$}
\bea
\phi_{0}(x,z)\phi_{\half}(0,0)&=&|z|^{-h_1-h_2+h_J}|x|^{-j_1-j_2+J}[C_J(0,0)\ln|x|+D_J(0,0) +... \\
\phi_{j_3}(x,z)\phi_{j_4}(0,0)&=&|z|^{-h_3-h_4+h_J}|x|^{-j_3-j_4+J}[E_J(0,0)\ln|x|+F_J(0,0) +... \nonumber 
\eea

where again  $J=j_1+\half=\half$. Then:
\bea
& &\la C_J^a E_J \ra = 0, \quad 
\la  C_J(x,z)F_J(0,0)\ra = \la  D_J(x,z)E_J(0,0)\ra = |z|^{-2h_J}|x|^{-2J}  \\ \nonumber
& &\la  D_J(x,z)F_J(0,0)\ra = -2|z|^{-2h_J}|x|^{-2J} \ln|x| \nonumber
\eea

We find these OPEs and the ones from our other solutions are consistent. In particular we find:
\be
\phi_j(x,z)\phi_0(0,0)  = \la \phi_j(x,z)\phi_0(0,0) \ra + \alpha E_j^x\ln|x| + \beta(2j-1)E_j^z\ln|z|+\gamma F_j+...
\ee

Also for $j_3,j_4 \ne 0$
\bea
\phi_{j_3}(x,z) \phi_{j_4}(0,0)=& & \la \phi_{j_3}(x,z) \phi_{j_4}(0,0) \ra + \\ \nonumber 
& & \quad \sum_{J}|z|^{-h_3-h_4+h_J}|x|^{-j_3-j_4+J}
\Bigl( C_{Jj_3j_4}O_J(0)  \\
& & \quad +~\delta_{J,\half}D_{j_3j_4}O^{'}_J(0)\ln|x| \Bigr)
\eea

The relative normalisation of the coefficients $C_{Jj_3j_4},D_{j_3j_4},\alpha,\beta,\gamma$ should be determined by the consistency of the  OPE with the $\SL$ Kac-Moody algebra. Although these OPEs are not complete they at least give us a hint of the possible structure and in particular some interesting behaviour at $j=\half$. In $\SL$ WZNW $j=\half$ is the pre-logarithmic operator \cite{Kogan:1997fd} that is the analogue of the puncture operator in Liouville. In $AdS_3$ it is also the critical point below which solutions in the bulk are not square integrable.

\end{itemize}

\section{Symmetries of the Correlation function}

The four point correlation function is fully determined by $F(x,z)$ and we now wish to comment on a few of the discrete symmetries of the $SL(2,R)$ model that are manifest in the KZ equation.

\subsection{Reflection $j\rightarrow 1-j$}

Under $j\rightarrow 1-j$ we find the Casimir $j(j-1)$ is unchanged and thus fields with the same other quantum numbers may be expected to mix. It is interesting that we can also see this symmetry from the point of view of the KZ equation. To do this we differentiate the KZ equation $1-2j_2$ times then we find that

\be \frac{\partial^{1-2j_2}}{\partial x^{1-2j_2}}F_{j_1,j_2,j_3,j_4}(x,z) \quad {and} \quad F_{j_1,1-j_2,j_3,j_4}(x,z) \label{eq:reflect}\ee satisfy the same KZ equation. This is due to the fact that, up to normalisation:
\be
\frac{\partial^{1-2j}}{\partial x^{1-2j}}\phi_j(x)=\phi_{1-j}(x)
\ee

This only makes sense if $1-2j \in Z_+$ which is exactly the case in which the series starting at $j$ and $1-j$ can be indecomposable.

\subsection{$j\rightarrow \frac{k}{2}-j$}

It can also be easily verified that:
\be (x-z)^{k-j_1-j_2-j_3-j_4}z^{\alpha}(1-z)^{\beta}F_{\frac{k}{2}-j_3,\frac{k}{2}-j_4,\frac{k}{2}-j_2,\frac{k}{2}-j_1} \ee
and $F_{j_1,j_2,j_3,j_4}(x,z)$
obey the same KZ with appropriately chosen $\alpha,\beta$. As was discussed in \cite{Maldacena:2000hw} precisely this symmetry is related to spectral flow.

\subsection{Strange $Z_2$ symmetry}

It is very easy to see that under the transformation:
\bea \label{eq:strange}
j_1&\rightarrow& \tilde{j_1}=\phantom{-}\frac{j_1}{2}+\frac{j_2}{2}-\frac{j_3}{2}+\frac{j_4}{2} \nonumber \\
j_2&\rightarrow& \tilde{j_2}=\phantom{-}\frac{j_1}{2}+\frac{j_2}{2}+\frac{j_3}{2}-\frac{j_4}{2} \nonumber \\
j_3&\rightarrow& \tilde{j_3}=-\frac{j_1}{2}+\frac{j_2}{2}+\frac{j_3}{2}+\frac{j_4}{2} \nonumber \\
j_4&\rightarrow& \tilde{j_4}=\phantom{-}\frac{j_1}{2}-\frac{j_2}{2}+\frac{j_3}{2}+\frac{j_4}{2}
\eea
$\CP$ and $\CQ$ shift by a constant. If we consider:
\be z^{\frac{\alpha}{k-2}}(z-1)^{\frac{\beta}{k-2}}F_{\tilde{j_1},\tilde{j_2},\tilde{j_3},\tilde{j_4}} \ee
where $\alpha=-\half(j_1-j_2)^2+\half(j_3-j_4)^2, ~~~ \beta=-\half(j_2-j_3)^2 +\half(j_1-j_4)^2$. 

Then we find that this satisfies exactly the same KZ equation as $F_{j_1,j_2,j_3,j_4}(x,z)$ and thus can, presumably by uniqueness of the solutions, be identified with it; at least up to some overall scale. This strange symmetry was noted earlier in \cite{Andreev:1995bj} in the Dotsenko-Fateev model \cite{Dotsenko:1984nm} for specific values of spin $J_0^3$. It was shown to follow from an integral identity in \cite{Petersen:1996hs} in the case of $SL(2)$ degenerate representations.

\section{String theory on $AdS_3$}

String theory in $AdS_3$, which is the $\SL$ group manifold, is described by the $\SL$ WZNW theory. There has been much early work discussing the spectrum and consistency of such a theory \cite{Gepner:1986wi,oldstring,PGB}. There has been renewed interest recently \cite{newstring} due to the AdS/CFT correspondence \cite{Maldacena:1998re}. The duality is between string theory in the bulk of Anti-de-Sitter space (AdS) and a conformal field theory (CFT) on the boundary of the space-time. Bulk fields naturally couple to local operators in the boundary CFT. Correlation functions on the boundary can be computed from the bulk theory in the supergravity approximation by taking the classical tree level graphs for the bulk interactions. A precise recipe to compute this is given in \cite{Gubser:1998bc,Witten:1998qj}. For the $AdS_5$ case for example see \cite{Muck:1998rr}.

So far most work has been conducted in this supergravity approximation. For general $AdS_n$, this is because we do not understand how to describe the full string theory in such a background. For the case of $AdS_3$ however the worldsheet theory is described by the $\SL$ WZNW model and the boundary theory is a two dimensional CFT and so there is the possibility to go beyond this leading supergravity approximation. Fields are naturally classified according to the representation theory of $\SL$. It is also an extremely interesting case in which we have an exactly solvable string theory and thus we can study physics in this background, for instance the issues of unitarity in such a curved space.

In \cite{Kogan:1996df} it was suggested that logarithmic operators in the worldsheet generate zero modes in the target space which restore the symmetries that are broken by the background. This was used in \cite{Periwal:1996pw} to describe D-brane recoil. For the latest developments in this area see \cite{Campbell-Smith:2000qv} and references therein. In the case of $AdS_3$ geometry, which is the near horizon limit of $D1-D5$ system, these zero modes should restore the full Poincare invariance broken by the position of the branes. It would be interesting to see if our solutions could be interpreted in this context.

Short distance logarithmic singularities in four point correlation functions in the boundary CFT came as a surprise. Short distances in boundary theory are mapped to long distances in the bulk and vice versa via the familiar UV/IR relation \cite{Susskind:1998dq}. In the $AdS_{5}/CFT_{4}$ case, in which the boundary theory is ${\mathcal{N}}=4$ supersymmetric Yang-Mills, these singularities are interpreted as an outcome of perturbative expansion in terms of anomalous dimensions of unprotected multi-trace operators which correspond to multiparticle states in the bulk \cite{D'Hoker:1999jp,Hoffmann:2000mx}. In this case it is expected that as the boundary theory in this case is reasonably well understood that these logarithms are not due to logarithmic structure in ${\mathcal{N}}=4$ SYM. This also seems to be the case in $AdS_{3}/CFT_{2}$ as it has been shown that the spectrum of multiparticle states in $AdS_{3}$ agrees precisely with the multi-trace operators in boundary CFT \cite{deBoer:1998ip}. However in $AdS_3$ the grey body spectrum has a logarithmic singularity which cannot be explained by an anomalous dimension. In \cite{Myung:1999nd} it was shown that this could not be due to a LCFT on the boundary with degenerate primary states. This paradox was solved in  \cite{Lewis:1999qv} which suggested that a LCFT with degeneracy between a primary state and the descendant of another could give the calculated spectrum. It is thus still an open question as to the role of LCFT in the boundary theory particularly beyond the supergravity approximation.

The $j=0$ operators on the world sheet correspond to the non-normalisable scalar singletons in the bulk of $AdS_3$. In general $AdS_n$ these are special finite dimensional representations which lie at the limit of unitarity \cite{Dirac:1963ta,Fronsdal:1989pp,HarunarRashid:1992bv}. It was shown in \cite{Kogan:1999bn} that if one considers singletons in the bulk then they would naturally induce a LCFT on the boundary with correlation functions given by (\ref{eq:logcorrel}). This form of the bulk action was also considered in \cite{Ghezelbash:1998rj} but from a different point of view. As we have found logarithmic correlation functions for $j=0$ operators on the worldsheet there might be a relation between the possible logarithmic structure of the worldsheet and boundary LCFTs.

\section{Conclusions and Discussion}
We have found several solutions exhibiting logarithmic behaviour in both the Kac-Moody and Virasoro parts. These clearly show that the set of primaries is not complete and that extra fields can be generated in the OPE. At the free level no-ghost theorems have been proved \cite{Evans:1998qu} for $0<j<\frac{k}{2}$, and null states are expected to decouple, but general statements about the fully interacting case cannot be made until the full field content has been established. There has been several attempts to compute correlation functions using other methods \cite{Gawedzki:1989rr,Teschner:1999ug,otherapp}. An analysis of two point functions \cite{otherapp} shows that $j$ should obey a more stringent bound $\half<j<\frac{k-1}{2}$ (See also \cite{Maldacena:2000hw} for a discussion of this bound from a different perspective). All the exact logarithmic solutions found by us involve some operators outside this range. Correlation functions involving primaries with $j<1/2$ can be mapped into correlation functions with $j>1/2$ using reflection symmetry (\ref{eq:reflect}) and logarithms disappear. The special case $j=\half$, which is at the boundary of discrete and continuous series, deserves further study \cite{Lewis:2000tn}. In \cite{Giribet:1999ft} the free field approximation was used, valid near to the boundary of $AdS_3$, and it was found that zero and negative norm states could be produced by interactions. It would be interesting to compare these different approaches. Our results seem to suggest that at the interacting level it is not possible to remove all the zero norm states as a logarithmic pair is created. 

There are many aspects which we feel deserve further attention. It is not obvious which indecomposable representations are allowed in the $\SL$ model. The KZ equation for a non-compact group, unlike the compact case, does not reduce to an ordinary differential equation and so there are issues of uniqueness and boundary conditions. The bulk degrees of freedom corresponding to the logarithmic operators are presumably singletons \cite{Kogan:1999bn} and it would be interesting to see how these enter in the bulk action. It is also clear that these are perhaps not the most general logarithms that could occur as the Casimir $J^2$ is always diagonal in our representation which is not required for indecomposable representations. We hope to study some of these in future work.

\section{Acknowledgements}

We would like to thank I. Kogan, A.M. Tsvelik, and M.J. Bhaseen for interesting and helpful discussion. We would also like to thank J. Rasmussen for drawing our attention to one of his papers which also discussed the strange symmetry. The research of A.N is funded by PPARC studentship number PPA/S/S/1998/02610. Sanjay is funded by the Felix Scholarship (University of Oxford).

\newpage

\appendix

\section{$\SL$ representation theory}
For completeness we include the representations of $\SL$ for the primary fields. The representations are characterised by the eigenvalues of the quadratic Casimir
\be \label{eq:casimir} C_2=\half(J_0^+ J_0^- + J_0^- J_0^+)-J_0^3 J_0^3= -j(j-1) \ee
and also the eigenvalue of $J_0^3$.  This set includes the unitary representations as well as the non-unitary ones and also importantly the non-decomposable representations as all of these could contribute to the operator product expansion. The fact that not all representations can be written in a block diagonal (decomposable) form is due to the fact that more than one field may possess the same quantum numbers $j$ and $m$ and so in general $J_0^3$ can only be reduced to Jordan normal form rather than fully diagonal.
The representations are all built starting from a primary field and then acting with $J_0^{\pm}$. They are

\begin{itemize}
\item Continuous representations: \\ 
This is an irreducible, infinite dimensional representation that
does not have a highest or lowest weight state.The $j$ and $m$ eigenvalues are unrelated. A representation of this type has the states
\be \left.\{|j,\alpha;m\right>: m=\alpha,\alpha\ \pm 1,\alpha\ \pm 2, \cdots \} \ee
where $J_0^3 \left. |j,\alpha;m\right>=m \left. |j,\alpha;m\right>$ and also $j\pm \alpha \notin \IZ $ so that the representation is unbounded in either direction. We shall only consider real valued of $\alpha$ and so without loss of generality we can take $0 \leq \alpha < 1$ . 
For fixed $C_2$ and $\alpha$,
the representations for each branch of  (\ref{eq:casimir})\ are equivalent. 
Imposing unitarity restricts us to two types of representations:
\begin{enumerate}
	\item ${\mathcal{C}}^{\alpha}_j$ --``Principal continuous series'' 	occurs for $C_2 < -\frac{1}{4}$;
	thus $j = \half + is$ with $s \in \IR$
	\item ${\mathcal{E}}^{\alpha}_j$--``Complementary series'' occurs
	for $C_2> - \frac{1}{4}$.  Here $j$ is real with $\half < j < 1$ and
	$j - \half < |\half - \alpha | $.
\end{enumerate}
\item Discrete lowest weight representation. \\
This is an irreducible, infinite-dimensional lowest
weight representation and exists for $2j \notin \IZ^-\cup\{0\}$. 
A representation of this type has the states
\be \left.\{|j;m\right>: m=j,j+1,j+2, \cdots \} \ee
where $J_0^-\left.\{|j;j\right>=0$ and $J_0^3 \left. |j;m\right>=m \left. |j;m\right>$

This representation will be unitary if
$j > 0$. For representations of the group $\SL$ then $2j \in \IZ^+$ but for the universal cover $j \in \IR^+$.

The derivation of this
representation in \cite{barut}\ shows that it can be embedded in a
reducible, nondecomposable representation where $m-j$ is an arbitrary
integer. We can act on these states with $m<j$ repeatedly using $J_0^+$ but once we get into the normal irreducible representation we cannot leave it with actions of $J_0^-$ because of the highest weight state.
\item Discrete highest weight representation: \\ 
This is an irreducible,  infinite-dimensional highest weight representation
and again exists for $2j \notin \IZ^- \cup\{0\}$.
A representation of this type has the states
\be \left.\{|j;m\right>: m=-j,-j-1,-j-2, \cdots \} \ee
where $J_0^+\left.\{|j;-j\right>=0$ and $J_0^3 \left. |j;m\right>=m \left. |j;m\right>$

This representation will be unitary if
$j > 0$. For representations of the group $\SL$ then $2j \in \IZ^+$ but for universal cover again $j \in \IR^+$.

In \cite{barut}\ this is embedded in a reducible, nondecomposable representation which contains $m>-j$ similar to that for the lowest weight representations.
\item Finite dimensional representations: \\
This is an irreducible, finite-dimensional representation.
It occurs when $2j \in \IZ^-\cup\{0\}$.
A representation of this type has the states
\be \left.\{|j;m\right>: m=|j|,|j|-1,|j|-2, \cdots -|j| \} \ee
where $J_0^-\left.\{|j;-|j|\right>=0$ , $J_0^+\left.\{|j;|j|\right>=0$ and $J_0^3 \left. |j;m\right>=m \left. |j;m\right>$.
The representation is only unitary in the case $j = 0$, i.e. for the 
identity representation, 
also known as the singleton~\cite{Dirac:1963ta,Fronsdal:1989pp,HarunarRashid:1992bv}.
It is contained
in a reducible nondecomposable representation for which $m$ is
arbitrary.

\end{itemize}

\section{Exact solutions of the KZ equation}

We can find particularly simple solutions to (\ref{eq:KZ})~ if the terms in $\CP$ and $\CQ$ with no derivatives in $x$ are zero. We thus look for solutions in which these terms vanish. However rather than do this naively we first make the substitution 
\be F(x,z)=z^{\frac{\alpha}{k-2}}(1-z)^{\frac{\beta}{k-2}}x^p(1-x)^qG(x,z) \ee

This is mostly redundant but does allow us to see the various crossing symmetries in the solutions.

The KZ equation (\ref{eq:KZ})~ then becomes
\be (k-2)\frac{\p G(x,z)}{\p z}=\frac{\tilde\CP G(x,z)}{z}+\frac{{\tilde\CQ } G(x,z)}{z-1} \label{eq:newKZ} \ee

We write $\tilde\CP =\tilde\CP _{\p}+\tilde\CP _0$ where $\tilde\CP _{\p}$ denotes the part of $\tilde\CP $ with derivative terms in $x$ and $\tilde\CP _0$ is the remainder. Of course $\tilde\CP _0$ is just an algebraic expression whereas $\tilde\CP _{\p}$ is a differential operator. Similarly we decompose $\tilde\CQ =\tilde\CQ _{\p}+\tilde\CQ _0$. Explicitly these are:
\bea
\tilde\CP _{\p} \!\!\!\!&=& \!\!\! -x^2(1-x)\frac{\p^2}{\p x^2}+x((j_1+j_2+j_3-j_4+1+2q)x-2(1-x)(j_2+p)-2j_1)\frac{\p}{\p x} \\
\tilde\CP _0 \!\!\!\! &=& \!\!\! x(2j_2+p+q)(j_1+j_2+j_3-j_4+p+q)-\alpha + \frac{q(j_1-j_2-j_3+j_4-q)}{1-x} + const \nonumber \\
\tilde\CQ _{\p} \!\!\!\! &=& \!\!\! -x(1-x)^2\frac{\p^2}{\p x^2}-(1-x)((j_1+j_2+j_3-j_4+1+2p)(1-x)-2x(j_2+q)-2j_3)\frac{\p}{\p x} \nonumber \\
\tilde\CQ _0 \!\!\!\!&=& \!\!\! -x(2j_2+p+q)(j_1+j_2+j_3-j_4+p+q)-\beta + \frac{p(-j_1-j_2+j_3+j_4-p)}{x} + const \nonumber
\eea

If we choose $\alpha$ and $\beta$ correctly to absorb the constant terms then we need not be concerned about these.

Although these expressions for general $j_i$ are quite complicated we can search for simple solutions using the ansatz $G(x,z)=X(x)+Z(z)$. Then

\be (k-2)\frac{dZ(z)}{dz}=
\frac{\tilde\CP _{\p}X(x)}{z}+\frac{\tilde\CP _0(X(x)+Z(z))}{z}+
\frac{\tilde\CQ _{\p}X(x)}{z-1}+\frac{\tilde\CQ _0(X(x)+Z(z))}{z-1} \ee

If we impose the vanishing of $\tilde\CP _0$ and $\tilde\CQ _0$ then we have:

\be (k-2)\frac{dZ(z)}{dz}=\frac{\tilde\CP _{\p}X(x)}{z}+\frac{\tilde\CQ _{\p}X(x)}{z-1} \ee As this is to be true for all $z$ we must have \be  \tilde\CP _{\p}X=r ~~~ \tilde\CQ _{\p}X=s \label{eq:PrQs} 
\ee

Thus if we seek solutions with this ansatz then we must impose the constraints $\tilde\CP _0=0$ and $\tilde\CQ _0=0$ and then find solutions $X(x)$ that satisfy (\ref{eq:PrQs}). Of course $G(x,z) \equiv {constant} $ is always a solution to (\ref{eq:newKZ}). In the next section we search for less trivial ones.

\subsection{Solution to the constraints $\tilde\CP _0=0$ and $\tilde\CQ _0=0$}

Solving the constraints we find the following cases:

\begin{center}
\( \begin{array}{ccccccc} 
Case & p & q  & j_1 & j_2 & j_3 & j_4   \\ 
\\
1 & 0 & 0 & -- & 0 & -- & --  \\
2 & 0 & -2j_2 & j_3\!-\!j_2\!-\!j_4 & -- & -- & --    \\
3 & 0 & -j_2\!-\!j_3\!+\!j_4 & 0 & -- & -- & --   \\
4 & 0 & 0 & -- & -- & j_4\!-\!j_1\!-\!j_2 & --  \\
5 & j_3\!-\!j_1\!-\!j_2 & j_1\!-\!j_2\!-\!j_3  & -- & -- & -- & 0   \\
6 & -2j_2 & 0  & -- & j_1\!-\!j_3\!-\!j_4 & -- & --  \\
7 & -j_1\!-\!j_2\!+\!j_4 & 0  & -- & -- & 0 & --  \\
8 & -2j_1 & -2j_3  & -- & -- & -- & j_2\!-\!j_1\!-\!j_3 \\
\end{array} \)
\end{center}

Also \be \alpha=p(1-2j_1-2j_2-p)+q(-j_1+j_2+j_3-j_4+q)-2j_1j_2 \ee
\be \beta=q(j_1+j_2-j_3-j_4+1+2p)+p(2j_1+4j_2-2j_4+2p)+2j_2(j_1+j_2-j_4)\ee

Although each of these gives different solutions for $F(x,z)$ we find that the actual physical quantity, the correlation function (\ref{eq:correl}), is the same in cases 1,3,5,7 and in 2,4,6,8 due to the prefactor multiplying $F(x,z)$ and thus these give us no new information. Therefore we concentrate on the first and fourth cases as these are simpler due to $p=q=0$ condition.

Thus we have 
\begin{itemize}
\item $j_2=0 ~~~ F(x,z)=X(x)+Z(z)$
\item $j_3=j_4-j_1-j_2 ~~~ F(x,z)=z^{\frac{-2j_1j_2}{k-2}}(1-z)^{\frac{-2j_2j_3}{k-2}}(X(x)+Z(z))$
\end{itemize}

\subsubsection{Case I:~~$j_2=0$}
\label{sec:case1}

Then we have:
\bea
\tilde\CP _\p X(x)\!\!\!\!&=&\!\!\!\!x^2(x-1)\frac{\p^2 X(x)}{\p x^2}  + \!((j_1\!+\!j_3\!-\!j_4\!+\!1)x^2-2j_1x)\frac{\p X(x)}{\p x} \\
\tilde\CQ _\p X(x)\!\!\!\!&=&\!\!\!\! -x(1-x)^2\frac{\p^2 X(x)}{\p x^2}  - \!((j_1\!+\!j_3\!-\!j_4\!+\! 1)(1-x)^2 \!- \! 2j_3(1-x))\frac{\p X(x)}{\p x}
\eea
Rather than immediately solving $\tilde\CP _\p X=r$,$\tilde\CP _\p X=s$ we note that
\bea
(x-1)\tilde\CP _\p X(x)+x\tilde\CQ _\p X(x)&=&r(x-1)+sx \\
\Rightarrow (j_1+j_3+j_4-1)x(1-x)\frac{\p X(x)}{\p x}&=&r(x-1)+sx  \eea
\par

If $j_1+j_3+j_4-1\neq 0$ then 
\be 
\label{eq:Xprime}
\frac{\p X(x)}{\p x}=\frac{1}{j_1+j_3+j_4-1} \left( \frac{r}{x}+\frac{s}{x-1} \right) \ee
It now remains to find if there are any extra constraints if we force this to be a solution of (\ref{eq:PrQs}). We find that for non-trivial solutions (i.e $r,s\neq 0$)

\begin{enumerate}
\item $j_1=j_3=j_4=0=j$
\item $r=-s ~~j= j_1=j_3\neq \half ~~ j_4=0$
\item $r=0 ~~ j=j_3=j_4\neq \half~~ j_1=0$ 
\item $s=0 ~~ j=j_1=j_4 \neq \half ~~ j_3=0$
\end{enumerate}

The case in which $j_i \equiv 0$ was the one previously found in \cite{Kogan:1999hz}. 

The solution of KZ equation in all these case follows from (\ref{eq:Xprime})  is of the form:
\be
F(x,z)=r \left[ \frac{\ln z}{k-2}+\frac{1}{2j-1}\ln x \right]+s \left[ \frac{\ln(z-1)}{k-2}+\frac{1}{2j-1}\ln(x-1) \right]+C
\ee
\\
If $j_1+j_3+j_4-1=0$ then we have $r=s=0$ and $F(x,z)$ has no $z$ dependence.
\bea
\tilde\CP _\p X(x)&=&0 \quad \tilde\CQ _\p X(x)=0 \\
\Rightarrow \frac{\p X(x)}{\p x}&=&(1-x)^{-2j_3}x^{-2j_1} \\
\label{eq:nologz}
\Rightarrow F(x,z)&=&\int (1-x)^{-2j_3}x^{-2j_1} dx
\eea
This has logarithmic behaviour in $x$ if $2j_1$ or $2j_3 \in Z^+$. However we see that the logarithmic behaviour in $z$ has vanished. If say $j_4=0$ this can be thought of as the limiting case of the above solutions when we reach the special values $j_1=j_3=\half$.

\subsubsection{Case II:~~ $j_3=j_4-j_1-j_2$}

We can repeat the same arguments to find non-trivial solutions. We find only the same solutions as before with logarithms in both $x$ and $z$. However we can find solutions with no logarithms in $z$ similar to (\ref{eq:nologz}) when $j_4=\half$:
\be
F_{j_1,\half-j_1-j_3,j_3,\half}=z^{-2j_1j_2}(1-z)^{-2j_2j_3}\int{x^{2j_3-1}(1-x)^{2j_1-1} dx}
\ee

A rearrangement of this solution is analysed in (\ref{eq:nojzero}). It is simpler however to observe that this is related to the previous case \ref{sec:case1} by the strange symmetry (\ref{eq:strange}).
If $j_4=j_2-j_1-j_3$ then we find this gives the same $F(x,z)$ as 
\bea
j_1&\rightarrow &j_1+j_2 \\
j_2& \rightarrow & 0 \\
j_3& \rightarrow & -j_1+j_4 \\
j_4& \rightarrow & -j_2+j_4
\eea

%\bibliographystyle{plain} % Alphabetical order
%\bibliographystyle{h-elsevier2} % Nuclear Physics format- no titles
%\bibliographystyle{unsrt} % In order of reference with titles
%\bibliography{ADSrefs,extras}
%\nocite{*}

\end{document}